\documentclass[manuscript]{aastex}

\newcommand{\msini}{M \sin i}
\newcommand{\mjup}{M_{\rm JUP}}
\newcommand{\mps}{{\rm ms}^{-1}}

\newcommand{\myemail}{harakawa@geo.titech.ac.jp}

\slugcomment{ApJ}

\shorttitle{New Exoplanet}
\shortauthors{Harakawa et al.}

\begin{document}

\title{Detection of a low-eccentricity and super-massive planet to the subgiant HD 38801\altaffilmark{1,2}}

\author{Hiroki Harakawa\altaffilmark{3}, Bun'ei Sato\altaffilmark{4}, Debra A. Fischer\altaffilmark{5,6}, 
Shigeru Ida\altaffilmark{3}, Masashi Omiya\altaffilmark{7}, John A. Johnson\altaffilmark{8}, 
Geoffrey W. Marcy\altaffilmark{9}, Eri Toyota\altaffilmark{10}, Yasunori Hori\altaffilmark{3}, 
and Andrew W. Howard\altaffilmark{9}}

\email{\myemail}

\altaffiltext{1}{Based on data collected at the Subaru Telescope, which is operated by the National Astronomical Observatory of Japan}
\altaffiltext{2}{Based on observations obtained at the W. M. Keck Observatory, which is operated by the University of 
California and the California Institute of Technology. Keck time has been granted by NOAO and NASA.}
\altaffiltext{3}{Department of Earth and Planetary Sciences, Tokyo Institute of Technology, 2-12-1 Ookayama, Meguro-ku, Tokyo 152-8551}
\altaffiltext{4}{Global Edge Institute, Tokyo Institute of Technology, 2-12-1-S6-6 Ookayama, Meguro-ku, Tokyo 152-8550}
\altaffiltext{5}{Department of Physics \& Astronomy, San Francisco State University, San Francisco, CA 94132}
\altaffiltext{6}{Department of Astronomy, Yale University, New Haven, CT 06511}
\altaffiltext{7}{Department of Physics, Tokai University, 1117 Kitakaname, Hiratsuka, Kanagawa 259-1292}
\altaffiltext{8}{Institute for Astronomy, University of Hawaii, Honolulu, HI 96822}
\altaffiltext{9}{Department of Astronomy, University of California, Berkeley, Berkeley, CA}
\altaffiltext{10}{Kobe Science Museum, 7-7-6 Minatojima-Nakamachi, Chuo-ku, Kobe, Hyogo 650-0046}


\begin{abstract}
We report the detection of a large mass planet orbiting around the K0 metal-rich subgiant HD38801 ($V=8.26$) by precise 
radial velocity (RV) measurements from the Subaru Telescope and the Keck Telescope. The star has a mass of $1.36M_{\odot}$ and metallicity of [Fe/H]= +0.26.
The RV variations are consistent with a circular orbit with a period of $696.0$ days and a velocity semiamplitude of $200.0\mps$, which yield a minimum-mass for the companion of $10.7\mjup$ and semimajor axis of $1.71$ AU.
Such super-massive objects with very low-eccentricities and hundreds of days period are uncommon among the ensemble of known exoplanets. 
\end{abstract}

\section{Introduction}

Since 1995, precise RV measurements have unveiled more than 400 extrasolar planets from surveys that include roughly 3000 nearby stars \citep[e.g.,][]{2007ARA&A..45..397U}.
The planets show a great diversity in their physical attributes: masses of the planetary companions range from $4M_{\oplus} (\sim 0.013\mjup)$ to larger than $13\mjup$, with semimajor axes of 0.02AU - 6AU and orbital eccentricities of 0 - 0.9.
A large number of planets have enabled us to discuss correlations between each orbital element and their masses in terms of planet formation and orbital evolution.
For example, the diagram of orbital eccentricity against semimajor axis shows that eccentricities of planets with semimajor axis larger than 0.1 AU are almost uniformly distributed from 0 to 1, suggesting orbital evolution due to gravitational interaction between planets, while those inside 0.1 AU tend to be damped below 0.1 due to tidal circularization.
The diagram of eccentricity against planet mass also shows another possible hint for planet formation that super-massive planets ($>5 \mjup$) tend to have relatively high eccentricity ($>0.3$), while less massive planets have a wide range of eccentricities.

As the number of planet surveys has enlarged, diversity of planet-host stars has also been explored.
Several surveys have focused on specified stellar properties, particularly stellar mass \citep{2005A&A...443L..15B, 2010PASP..122..156A, 2008PASJ...60.1317S, 2007ApJ...665..785J} and metallicity \citep{2005ApJ...620..481F,2009ApJ...697..544S}, in order to investigate correlations between the stellar properties and planetary parameters such as stellar mass and occurrence rate of planets.
As a result from these surveys, the statistical correlations between stellar properties and planet parameters have begun to be unveiled.
For example, occurrence rate of Jovian planets increases with stellar metallicity from 3\% for stars with [Fe/H] $< 0$ to 25\% for those with [Fe/H] $> +0.3$ \citep{2005ApJ...620..481F}, and also increases with stellar mass from 2\% around M dwarfs ($< 0.6 M_{\odot}$) to approximately 9\% around F and A stars \citep[1.2 - 1.9 $M_{\odot}$;][]{2007ApJ...665..785J}.
Planetary mass could also be positively correlated with stellar metallicity and mass \citep{2007A&A...472..657L}.
Impacts of stellar properties on characteristics of planets, such as orbital eccentricity and multiplicity, are also expected to emerge as the number of planets grows.

The N2K program is a large scale international exoplanet-search project started in 2004,
which consists of a sample of 2000 metal-rich solar-type stars \citep{2005ApJ...622.1102F}.
The search originally targeted short-period planets with a high-cadence observational strategy.
However, it has also detected intermediate-period ($18 - 3810$ days) ones thanks to the long-term observations.
From the collective N2K surveys, we have discovered 22 planets so far, which have a wide variety of mass (0.22 - 13.1 $\mjup$) and orbital parameters ($a=$ 0.04 - 4.9AU, $e=$ 0 - 0.7), and orbit around a variety of host stars \citep[$-0.11 < {\rm [Fe/H]} < 0.37, 0.88 < M/M_{\odot} < 1.31$;][]{2007ApJ...657..533W,2006ApJ...647..600J,2009PASP..121..613P}.
The planets discovered from our surveys can thus contribute to our general understanding of planet formation and evolution depending on stellar properties.

Here we report the detection of an exoplanet orbiting the metal-rich solar type subgiant HD38801 discovered from the N2K sample at the Subaru Telescope and Keck observatory. 
The planet has a minimum mass of $10\mjup$ and a low eccentricity.
Such a planet has been rarely discovered so far around solar-type dwarfs and subgiants.
We describe the characteristics of HD38801 in \S \ref{sec_str}. In \S \ref{sec_obs}, we present our observations, and the orbit of HD38801b. We summarize our results in \S \ref{sec_sum} and discuss formation scenarios of such a low-eccentricity super-massive planet.

\section{Stellar characteristics}
\label{sec_str}
HD38801 (HIP 27384) is listed in the Hipparcos catalogue (ESA 1997) as a K0IV star with a visual magnitude 
of $V=8.26$, and a color index $B-V=0.873$. The {\it Hipparcos} parallax of 10.06 mas corresponds to a distance 
of 99.4 pc which yields the absolute visual magnitude of $M_V = 3.27$. A high-resolution spectroscopic analysis 
described in \cite{2005yCat..21590141V} derived an effective temperature $T_{eff} = 5222\pm44$ K, a surface gravity $\log g = 3.84\pm0.1$, rotational velocity $v\sin i = 0.54\pm0.5 {\rm km\ s}^{-1}$, and metallicity [Fe/H] = $0.26\pm0.03$ dex for the star.
The bolometric luminosity of $L_{star} = 4.56\pm0.47L_{\odot}$ was derived from the $M_V$ and a bolometric correction of $-0.195$ \citep{1996ApJ...469..355F}.
From the bolometric luminosity, parallax and $T_{eff}$, we derived a stellar radius of $2.53\pm0.13R_{\odot}$ based on the Stefan-Boltzmann relation. 

In order to estimate a stellar mass, we interpolated the metallicity, effective temperature, and 
luminosity onto the stellar interior model grids \citep{2002A&A...391..195G}.
We adopted the three-dimensional interpolation method described by \cite{2007ApJ...665..785J}, and applied 
7\% uncertainty based on a comparison among different stellar model grids.
Using the SME-derived stellar parameters, we estimated mass of $1.36\pm0.09M_{\odot}$ and an age of $4.67\pm2.56$ Gyr for HD38801.
We measured $S_{HK}$ of $-0.174$, which is core emission in the Ca II HK lines relative to the continuum, 
to obtain chromospheric activity. We derived $\log R'_{HK}=-5.0$, the ratio of flux from $S_{HK}$ to the 
bolometric stellar flux, which indicates that the star is chromospherically inactive.
However, it is known that subgiants show "jitter" of 4-6 $\mps$, which is velocity scatter in excess of 
internal errors due to astrophysical sources such as pulsation and rotational modulation of surface features, 
even with such a low value of $\log R'_{HK}$ \citep{2007ApJ...665..785J}.
We therefore adopt a jitter value of $6\ \mps$ which is added in quadrature 
($\sigma_{error}^2 = \sigma_{obs}^2 + \sigma_{jitter}^2$) to the formal 
velocity uncertainties for HD38801 when we fit a Keplerian orbit 
to the RV data.

\section{Doppler observations and Keplerian-fit}
\label{sec_obs}
The Subaru Telescope's component of the N2K survey uses the High Dispersion Spectrograph (HDS) on the 8.2-m Subaru Telescope \citep{2002PASJ...54..855N}.
We adopted the setup of StdI2b, which simultaneously covers a wavelength region of 3500-6100\AA\ by a mosaic of two CCDs.
The slit width was set to $0.8"$ for the first 3 observations (2005) and $0.6"$ for the last 15 data, giving a reciprocal resolution ($\lambda /\Delta \lambda$) of 45000 and 55000, respectively.
Using this setup, we can obtain a signal-to-noise ratio of S/N $\sim 150\ {\rm pixel}^{-1}$ at 
5500\AA\ with 110-300 sec exposure for our typical targets, $V \sim 8.5$ stars.
We used an iodine absorption cell to provide a fiducial wavelength reference for precise RV measurements \citep{2002PASJ...54..865K,2002PASJ...54..873S}. 
Although the cell was originally installed just behind the entrance of the spectrograph, it was moved in front of it in 2006.
We used an analysis code of \cite{2002PASJ...54..873S} for modeling iodine superposed spectra and deriving stellar RVs, which is based on the standard method developed by \cite{1996PASP..108..500B} and \cite{1995PASP..107..966V}.
Using the code and above spectrograph setting, we can achieve a Doppler precision of 4-5 $\mps$.

Between Sep 2006 and Jan 2009, we obtained 10 RV observations from the Keck 10-m telescope with the HIRES spectrograph \citep{1994SPIE.2198..362V}.
An iodine cell was placed in the optical path to provide both the precise wavelength solution and the PSF (Point Spread Function) of spectrograph instruments 
\citep{1992PASP..104..270M,1996PASP..108..500B}. We used the B5 decker corresponding to a slit width 
of 0.86" and a wavelength resolution of 65,000. Appropriate exposure times between 150 and 240 seconds 
were taken and this results the S/N of 150 for respective observations. The typical precision of recent 
Keck observation is 1.5 $\mps$.

\begin{figure}
\epsscale{1.0}
\plotone{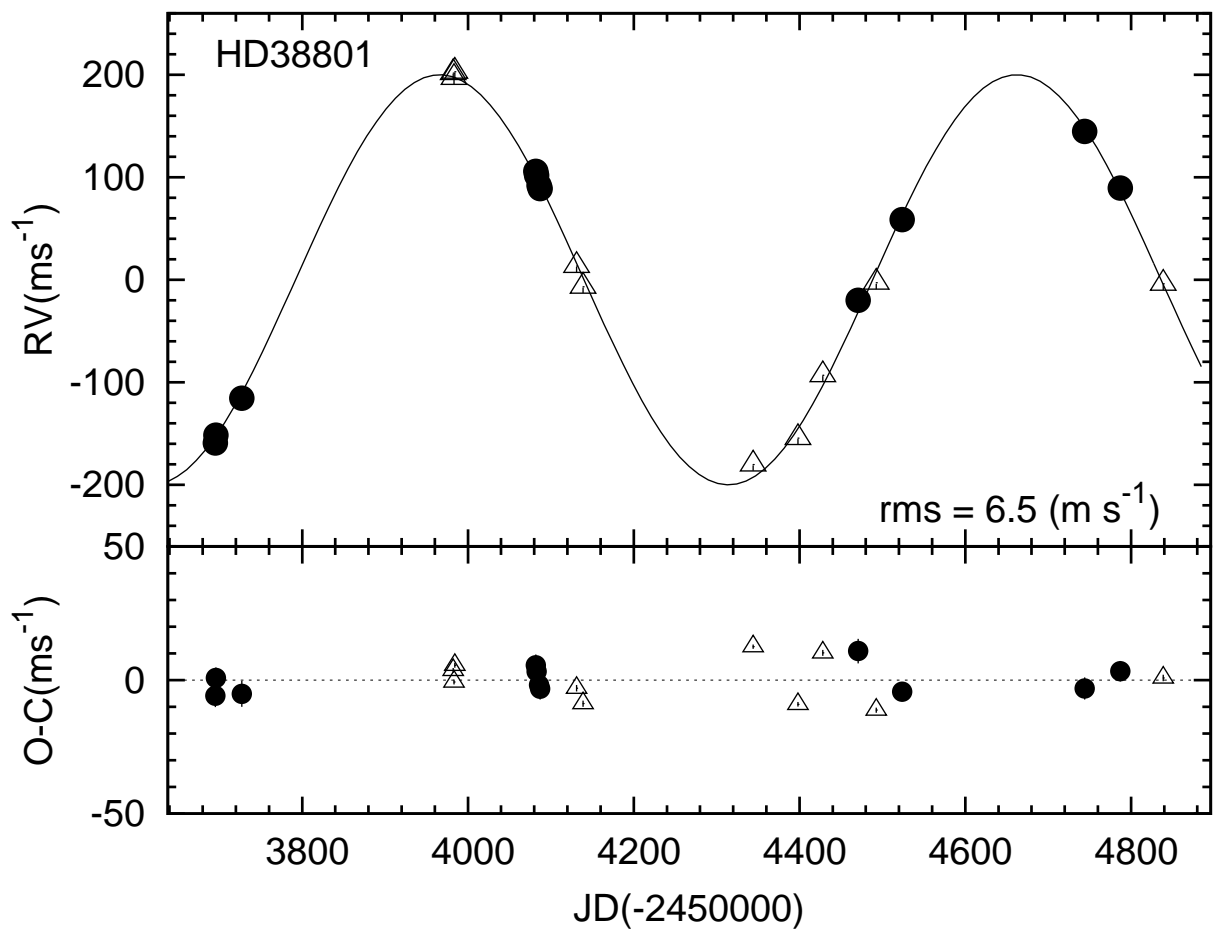}
\caption{The RVs for HD38801 derived from the Subaru Telescope({\it filled circles}) and the Keck Telescope({\it opened triangles}). With an orbital period of $696$ d, velocity semiamplitude of $200 {\rm ms}^{-1}$, and stellar mass of $1.36M_{\odot}$, the indicating planet mass is $M\sin i=10.7M_J$, the semimajor axis is $1.7 {\rm AU}$.
The eccentricity is fixed to 0.
The rms to the fit is $6.5{\rm ms^{-1}}$. The offsets of $-58.8{\rm ms}^{-1}$ and $20.5{\rm ms}^{-1}$ were applied to the Subaru RVs and the Keck RVs respectively.\label{fig_fit}}
\end{figure}

We made the first observations for HD38801 at the Subaru Telescope and detected large RV variations.
After that, the star has been monitored at the Subaru Telescope and the Keck Telescope for 3 years (from November 2005 to November 2008) 
and a total of 20 data points were obtained. The observation dates, RVs, and instrumental uncertainties 
are listed in Table \ref{tbl_obs} and plotted in Figure \ref{fig_fit}.

To derive the best-fit Keplerian orbit to the RV data, we adopted the down-hill simplex algorithm 
\citep[AMOEBA;][]{1965CJ..7.308} to reduce $\chi^2$ which has multiple parameters (orbital period $P$, 
time of periastron passage $T_c$, eccentricity $e$, velocity semiamplitude $K_1$, and argument of 
periastron $\omega$) and obtained the best fit parameters of single Keplerian orbit.
The stellar jitter of 6 $\mps$ was added in quadrature to the measurement uncertainties.
At first fit, when all of these parameters were let free, we obtained very low eccentricity of 0.04 and 
poorly constrained $\omega$. So we made the another fitting with the fixed values of $e$ and $\omega$ to 0.
The best fit parameters thus obtained are listed in Table \ref{tbl_prm}, and the Keplerian fit is overplotted 
on the RV data in Figure \ref{fig_fit}. We obtained a period of $P=696.3\pm2.7$ d and a velocity semiamplitude 
of $200.0\pm3.9\ \mps$. The rms scatter to this fit is 6.5 $\mps$ and 
$\left( \chi_{\nu}^2\right)^{1/2}=1.2$ as a reduced $\chi$. The uncertainties in the orbital parameters 
were estimated by a bootstrap Monte Carlo approach. The RV residuals from best-fit Keplerian curve are 
scrambled and added back to the original measurements, and then re-fit and derive new parameters.
Iterating 10,000 times, we obtained $1\sigma$ value of each parameters.
With a stellar mass of $1.36\pm0.09M_{\odot}$, we obtained a minimum-mass for the companion of 
$\msini = 10.7\pm0.5\mjup$ and semimajor axis of $a_{p} = 1.70\pm0.03 {\rm AU}$.

As seen in Figure \ref{fig_fit}, the residuals to the Keplerian fit in JD2454300-2454500 exhibit larger scatter than other parts.
We performed a periodogram analysis \citep{1982ApJ...263..835S} to the residuals.
However, we found no significant peaks in the periodogram at this stage.
More dense sampling of data will help verify the existence of any periodicity in the residuals and its origin.

\section{Summary and discussion}
\label{sec_sum}
We here reported the detection of a large mass planet in an almost circular orbit around a high metallicity K0 IV type star HD38801 from the precise RV observations at the Subaru Telescope and the Keck Telescope.
The star has a mass of $1.36M_{\odot}$ and metallicity of [Fe/H]=0.26.
The planet is a new sample around a metal-rich and relatively high-mass star, which makes an important role to investigate dependence of planets on host star's properties.

One remarkable feature of this planet is its low orbital eccentricity ($e\sim 0$) despite of the large mass ($\msini =10.7\mjup$) and the intermediate orbital distance ($a=1.7$AU). 
Such a planet with low eccentricity ($e<0.1$) and large-mass ($\msini > 10\mjup$) has been rarely discovered so far in the regions of $0.1\ {\rm AU}<a<5\ {\rm AU}$ around FGK dwarfs and subgiants.
Only HD16760b ($\msini=13\mjup, a=1.08\ {\rm AU}, e=0.084$), which was recently discovered from the Subaru Telescope and the Keck Telescope \citep{2009A&A...505..853B, 2009ApJ...703..671S}, falls into this category except for HD38801b.
In the core-accretion model, massive companions are generally formed with circular orbits because of protoplanetary disk's circular motion. The circular orbit of HD38801b is consistent with this scenario.
Tidal interaction between the host star and the planet cannot explain the circular orbit because the stellar radius is only $2.5R_{\odot}$ ($\sim0.01$AU), which is too small to circularize the planetary orbit
at 1.7 AU.
Contrary to HD 38801b, most of the intermediate-period ($>18$ days) and large-mass ($>5\mjup$) planets discovered around FGK dwarfs and subgiants so far tend to reside in eccentric orbits.
The origin of such planets still remains to be solved. It is generally thought that eccentric planets are formed by planet-planet scattering in multiple planetary systems.
The scenario suggests formation of multi super-massive planets in a system and predicts existence of scattered outer companions as massive as inner ones, which can be good targets for direct imaging.
Other scenarios such as giant impact\citep{2008A&A...482..315B},and disk instability\citep{1998ApJ...503..923B} and Kozai-mechanism when they are in binary systems \citep{1962AJ.....67..591K} have also been proposed to explain for formation of such massive eccentric planets.
Interestingly, planets with similar properties to HD38801b have been discovered around intermediate-mass giants.
Comparing populations of such planets between different types of stars would provide a hint on formation scenario and evolution for the planets.

The high-metallicity and the evolutionary stage of subgiant for HD38801 is another interesting feature of this system.
It is well known that the detection rate of giant planets shows positive correlation with the metallicity of the host stars \citep{2005ApJ...622.1102F}.
Two scenarios have been proposed to explain the origin of the correlation; one is that more metal-rich stars tend to form more giant planets and the other is that the stellar surface is polluted by planetary debris \citep[e.g.,][]{2004ApJ...616..567I}.
At the early stage of the main sequence, a stellar surface could be polluted due to the accretion of planetesimals and protoplanets from debris disk and metallicty of the stellar surface could be enhanced.
However, if the star has deep convective envelope, the metal-enhanced surface can be mixed with original stellar material and diluted by the convective flow.
Thus, the existence of a giant planet around metal-rich evolved star as the subgiant HD38801, which 
probably has deep convective envelope, favors the former scenario, that is giant planets tend 
to form in metal-rich environment \citep{2005ApJ...622.1102F}.
It should be noted, however, that many planets have been discovered even in metal-poor (${\rm [Fe/H]}<0$) giants, more evolved stars than subgiants, and no significant planet-metallicity correlation can be seen among them.
\cite{2007A&A...473..979P} proposed that the lack of metal-rich tendency is due to dilution by 
deeper convective envelope of giants than that of dwarfs and subgiants.
To make sure the origin of the high metallicity of planet-harboring dwarfs, it is important to increase the number of planets discovered around evolved stars with $<1.5M_{\odot}$, for which we can detect planets in all of the three evolutionary stages, dwarfs, subgiants, and giants, by precise Doppler technique.

We extend our gratitude to Akito Tajitsu and Tae-Soo Pyo for their great expertise and support 
of HDS observations from the Subaru Telescope.
We also gratefully acknowledge the efforts and dedication of all Keck Observatory staff.
This research has made use of the Simbad database, operated at CDS, Strasbourg, France.
We thank all Hawaiian people for their hospitality and native Hawaiian ancestry on whose 
sacred mountain of Mauna Kea we are privileged to be guests.
Without their generous supports, we would not be able to publish our new planet.

\clearpage

\begin{deluxetable}{lc}
\tabletypesize{\scriptsize}
\tablecaption{Stellar Parameters for HD 38801 \label{tbl_str}}
\tablewidth{0pt}
\tablehead{
\colhead{Parameter} & \colhead{Value(error)}
}
\startdata
$V$ &8.26\\
$M_V$ &3.27\\
$B-V$ &0.873\\
Spectral type &K0 IV(subgiant)\\
Distance(pc) &99.4\\
$T_{eff}$(K) &$5222(44)$\\
$\log g$ &$3.84(0.1)$\\
$\left[ {\rm Fe/H}\right]$ &$0.26(0.03)$\\
$v\sin i ({\rm km s^{-1}})$ &$0.54(0.5)$\\
$M_* (M_\odot )$ &$1.36(0.09)$\\
$R_* (R_\odot )$ &$2.53(0.13)$\\
$L_* (L_\odot )$ &$4.56(0.47)$\\
age(Gyr) &4.67(2.56)\\
$S_{HK}$ &$-0.174$\\
$\log R'_{HK}$ &$-5.01$
\enddata

\end{deluxetable}

\begin{deluxetable}{lccc}
\tabletypesize{\scriptsize}
\tablecaption{RVs for HD38801.\label{tbl_obs}}
\tablewidth{0pt}
\tablehead{
\colhead{Observation Date} & \colhead{RV} & \colhead{Uncertainties} & \colhead{\ } \\
\colhead{(JD $- 2,450,000$)} & \colhead{(${\rm ms^{-1}}$)} & \colhead{(${\rm ms^{-1}}$)} & \colhead{Observatory}
}
\startdata
3695.0902 &$-159.0$ &4.2 &Subaru\\
3695.9883 &$-151.3$ &4.0 &Subaru\\
3727.0304 &$-115.4$ &4.8 &Subaru\\
3982.1185 &201.4 &0.9 &Keck\\
3983.1122 &196.8 &0.9 &Keck\\
3984.1088 &202.8 &0.9 &Keck\\
4081.9436 &105.9 &4.0 &Subaru\\
4082.9788 &101.9 &4.7 &Subaru\\
4085.9284 &92.2 &4.8 &Subaru\\
4087.0406 &89.0 &4.1 &Subaru\\
4130.8895 &13.4 &1.2 &Keck\\
4138.9172 &$-6.9$ &1.2&Keck\\
4344.1185 &$-180.3$ &0.8 &Keck\\
4398.0809 &$-154.4$ &0.9 &Keck\\
4427.9607 &$-93.3$ &1.1 &Keck\\
4470.9194 &$-19.9$ &4.6 &Subaru\\
4492.7365 &$-3.1$ &1.0 &Keck\\
4523.7729 &59.0 &3.5 &Subaru\\
4744.0407 &145.0 &4.1 &Subaru\\
4787.0288 &89.6 &3.9 &Subaru\\
4838.9756 &4.0 &1.2 &Keck\\
\enddata
\end{deluxetable}

\begin{deluxetable}{lc}
\tabletypesize{\scriptsize}
\tablecaption{Spectroscopic Orbital Solution for HD38801b \label{tbl_prm}}
\tablewidth{0pt}
\tablehead{
\colhead{Parameter} & \colhead{Value(error)}
}
\startdata
$P$(d) &$696.3(2.7)$\\
$T_c$(JD) &$2,453,966.0(2.1)$\\
Eccentricity &0(fixed)\\
$\omega$ &0(fixed)\\
$K_1$($\mps$) &$200.0(3.9)$\\
$a_1\sin i$($10^{-3}$AU) &13(0.27)\\
$a_p$(AU) &$1.70(0.037)$\\
$\msini$($\mjup$) &$10.7(0.5)$\\
$N_{\rm obs}$ &21\\
rms($\mps$) &6.5\\
Reduced$(\chi_\nu^2)^{1/2}$ &1.2\\
\enddata

\end{deluxetable}

\clearpage

\clearpage
\begin{figure}
\epsscale{1.0}
\plotone{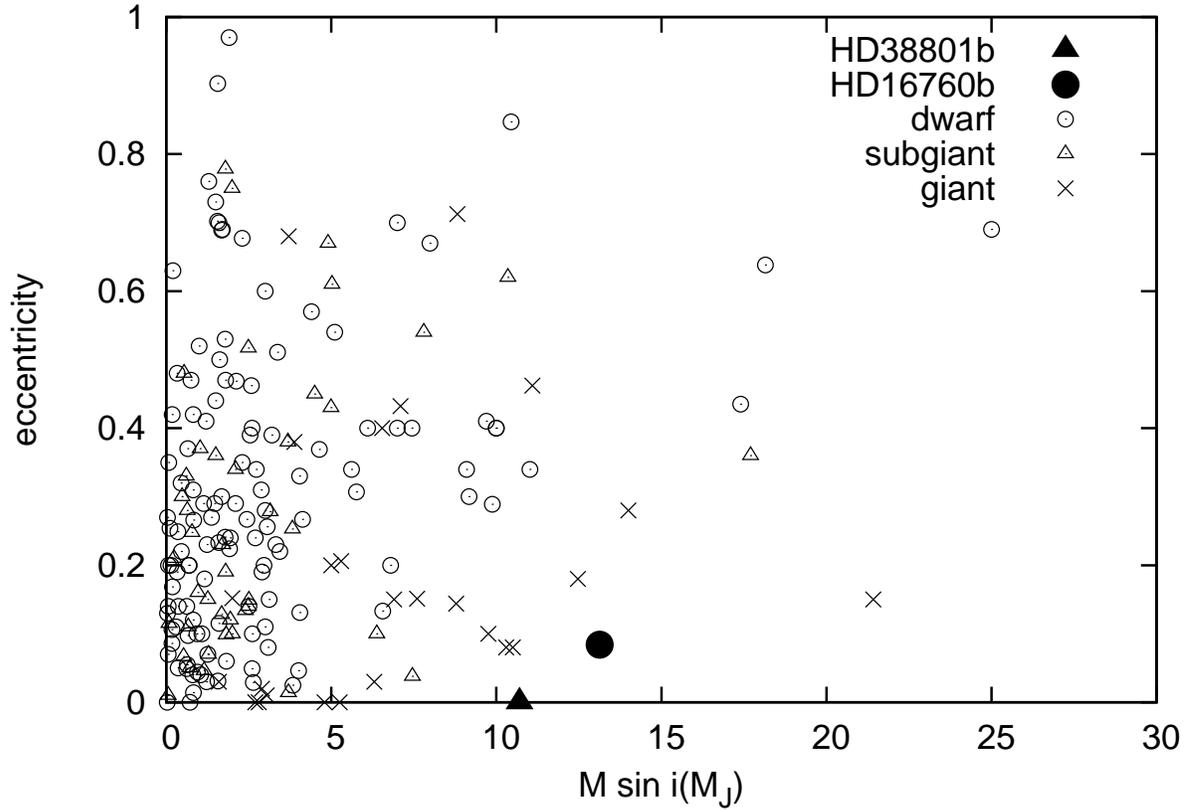}
\caption{The correlation diagram of the eccentricity and the minimum mass of the planets with above 14 days orbit.
HD38801b is plotted with {\it filled triangle}.
HD16760b({\it filled circle}), main sequence stars({\it opened circles}), and subgiants({\it opened triangles}), and giants({\it crosses}) are also plotted in.
The companion's properties of HD16760 is referred from \cite{2009ApJ...703..671S}
---http://exoplanet.eu \label{fig_m_e}}
\end{figure}

\end{document}